\documentclass[aps,pre,floats,twocolumn,showpacs,superscriptaddress]{revtex4}

\usepackage{graphicx,epsfig}
\usepackage{times}
\usepackage{graphics,dcolumn,bm,fleqn,epic,eepic,float}
\usepackage{amssymb,amsmath,multirow,rotate,color}
\begin{document}

\title{Local Empathy provides Global Minimization of Congestion in
  Communication Networks}

\author{Sandro Meloni}
\affiliation{Dipartimento di Informatica e Automazione, Universit\'a degli studi
  Roma Tre, Via della Vasca Navale, 79 00146, Roma, Italy}

\author{Jes\'us G\'omez-Garde\~nes}     
\affiliation{Departamento de Matem\'atica Aplicada, Universidad Rey Juan 
Carlos (ESCET), 95123 M\'ostoles (Madrid), Spain}       

\affiliation{Institute for Biocomputation and Physics of Complex
Systems (BIFI), University of Zaragoza, 50009 Zaragoza, Spain}

\date{\today}

\begin{abstract}
 We present a novel mechanism to avoid congestion in complex networks
 based on a local knowledge of traffic conditions and the ability of
 routers to self-coordinate their dynamical behavior. In particular,
 routers make use of local information about traffic conditions to
 either reject or accept information packets from their neighbors. We
 show that when nodes are only aware of their own congestion state
 they self-organize into a hierarchical configuration that delays
 remarkably the onset of congestion although leading to a sharp
 first-order like congestion transition. We also consider the case
 when nodes are aware of the congestion state of their neighbors. In
 this case, we show that empathy between nodes is strongly beneficial
 to the overall performance of the system and it is possible to
 achieve larger values for the critical load together with a smooth,
 second-order like, transition. Finally, we show how local empathy
 minimize the impact of congestion as much as global
 minimization. Therefore, here we present an outstanding example of
 how local dynamical rules can optimize the system's functioning up to
 the levels reached using global knowledge.
\end{abstract}

\maketitle


\section{Introduction}

Complex communication networks have recently attracted a lot of
attention from scientists due to the discovery of the topological
features of real communication systems such as the Internet
\cite{bookvesp}. The structure of these communication systems is
efficiently described by a graph in which nodes represent routers and
edges account for the communication channels. However, the structure
of these graphs is far from being purely random. Quite on the
contrary, they typically show a scale-free (SF) distribution for the
number of communication channels departing from and arriving to a
system's element.  The use of modern complex network theory
\cite{rev:albert,rev:newman,rev:bocc} together with tools inherited
from non-equilibrium statistical physics \cite{rev:doro2} have allowed
to study the dynamical properties of such communication systems. In
particular, this approach has been successfully applied to the study
of the structural evolution \cite{str1,str2} of the Internet, its
navigability \cite{nav1,nav2,nav3} and dynamical properties
\cite{flow0,flow1,flow2}, or the design of an efficient Digital Immune
System \cite{targ,holme,cohen,cov1,cov2,chen}.

A lot of the recent literature on communication networks has tackled
the critical properties of their jamming and congestion transitions
\cite{Ohira,Sole1,Sole2,cong1,Guimera,cong01,cong2,cong55,cong7,cong77}. These
studies have focused on the design of efficient routing strategies
that, on one hand, provide with short delivery times and, on the other
hand, avoid the onset of the congested state in which the load of
packets in the system increases, thus causing the failure of
information flow. It has been shown that finding the best suited
strategy depends strongly on two main features: the topological
patterns of the particular network and the load of information on top
of it. Regarding the first of these two issues, a number of routing
mechanisms have been studied on different structures
\cite{cong5,cong6,cong7,cong15} allowing to design resilient network
backbones \cite{cong1114,cong14,cong114}.

Many of the routing policies proposed so far rely on the (static)
structural properties of the communication network. Examples of such
policies are biased random walks \cite{cong888,cong88}, shortest-path
\cite{SP,cong8} and efficient-path \cite{EP} schemes.  These routing
mechanisms can be conveniently reformulated to incorporate the
information about the dynamical state of the system, {\em i.e.} the
congestion state of routers. This allows to dynamically change the
paths followed by information packets in order to bypass those
over-congested routes. In this line, congestion-aware schemes have
significantly improved the performance of biased random walks
\cite{DRW}, shortest-path \cite{cong3,cong4} and efficient-path
\cite{cong16} routings. In addition to the design of efficient routing
protocols, several strategies to avoid congestion have been
implemented. Remarkable examples of these strategies are the
implementation of incoming flow rejection \cite{cong10,cong11} and
packet-dropping mechanisms \cite{cong111} for avoiding the congestion
of single nodes, or the addition of a router memory to avoid packets
getting trapped between two adjacent nodes \cite{cong9}.

All the above studies have assumed that both network topology and the
mechanisms to avoid congestion are static ({\em i.e.} neither topology
nor the routing strategies change). However, this approach neglects
that, even for the same graph, the optimal routing policy depends
strongly on the state of congestion of the system
\cite{cong3,cong4,cong10,cong11,cong12,cong13}. Therefore, in order to
balance correctly the congestion in a communication system it seems
appropriate to allow the elements (routers) to switch to the best
suited strategy to avoid congestion given the instant traffic
conditions.  In this article, we propose an adaptive mechanism that
allows nodes to choose their individual strategies instead of imposing
a common policy. In this adaptive protocol routers exploit their local
information about the congestion state of the system to decide whether
to accept incoming packets. First, in section \ref{sec:minimal}, we
introduce a minimal routing model without any adaptive mechanism that
allows us to unveil the role of rejection when it is externally tuned.
In section \ref{sec:theo} we will consider that each router can adopt
its own rejection strategy and make some analytical derivations about
the optimal strategic configuration to avoid the congestion onset. In
section \ref{sec:myopic} we will implement our first adaptive
mechanism and show that when nodes are allowed to dynamically adapt
their own strategy, while only being aware of their own congestion
state (myopic case), the onset of congestion is shifted to a larger
critical load (with respect to the static algorithm introduced in
section \ref{sec:minimal}). This improvement is due to the
self-organization of the strategies of nodes into degree-correlated
configurations. However, we will show that the delay of the onset of
congestion comes together with a sharp, first-order like, transition
that provides no dynamical signals about the onset of
congestion. Finally, in section \ref{sec:empathetic} we show that when
nodes are aware of the congestion state of its nearest neighbors and
empathize with them, it is possible to recover the former large
critical load together with a smooth phase transition, avoiding the
uncertain scenario of the myopic adaptive model. More importantly, we
will show that tuning conveniently the degree of empathy between
routers it is possible to recover, through a local mechanism, both the
congestion levels and the rejection patterns provided by the global
minimization introduced in section \ref{sec:theo}.


\section{Minimal traffic model} 
\label{sec:minimal}

Let us start by introducing the minimal traffic model in which the
adaptive algorithm will be implemented below. In this model, we
consider the transfer of information packets between adjacent routers
as a probabilistic event. Inspired by \cite{cong10,cong11}, we
consider a set of stochastic equations for describing the time
evolution of the queue length of the nodes at some time $t$, ${\bf
  q^t}=\{q_{i}^t\}$. The queue length of a given node, $q_{i}^t$, can
either increase or decrease due to several events. First, at each time
step and with probability $p$, a new packet is generated being added to
the queue of the node. Second, at each time step each node tries to
send a packet in its queue to any of its first neighbors. This packet
can be rejected by the chosen neighbor with some probability
$\eta$. If the packet is accepted, it may be removed from the system
with certain probability $\mu$. These two latter events mimic the
effects, although with some important differences, of an active queue
control strategy as the random early detection (RED) \cite{Floyd1}
present on Internet routers and the arrival of the packet to its final
destination respectively. Following the above ingredients we can write
the time-discrete Markov chain of the minimal traffic model as:
\begin{eqnarray}
q_{i}^{t+1}=q_{i}^t+p&+&\sum_{j=1}^{N}\frac{\Theta(q_j^t)A_{ji}}{k_{j}}(1-\mu)(1-\eta)\nonumber
\\
&-&\Theta(q_{i}^t)\sum_{j=1}^{N}\frac{A_{ij}}{k_{i}}(1-\eta)\;,
\label{nullmodel}
\end{eqnarray}
where $A_{ij}$ represents the ($i$, $j$) term of the adjacency matrix
of the network substrate and $\Theta(x)$ is the Heaviside step
function ($\Theta(x)=1$ if $x>0$ and $\Theta(x)=0$ otherwise). Since
our network is undirected and unweighted, the adjacency matrix is
defined as $A_{ij}=A_{ji}=1$ if nodes $i$ and $j$ are connected and
$A_{ij}=A_{ji}=0$ otherwise. The quantity $k_{i}$ is the degree of a
node $i$ ($\sum_{j}A_{ij}=k_{i}$), {\em i.e.} the number of routers
connected to it. The right-hand-side of equation (\ref{nullmodel})
contains two terms accounting for the incoming flow of packets that
arrive to the queue of node $i$, namely, $p$ (accounting for the
external load of packets) and the first sum (accounting for the
arrival of packets from its first neighbors). On the other hand, the
second sum in equation (\ref{nullmodel}) accounts for the probability
that a packet from $i$ is delivered to a first neighbor.

\begin{figure}[t!]
\epsfig{file=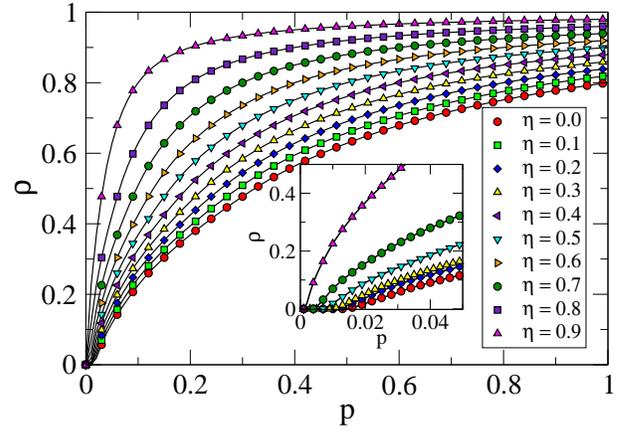,width=3.4in,angle=0,clip=1}
\caption{(Color online). Phase diagrams, $\rho(p)$, of the minimal
  traffic model using different values of the rejection rate
  $\eta$. The inset shows the existence of different critical values
  $p_c$ when varying $\eta$}
\label{fig:1}
\end{figure}

The set of equations (\ref{nullmodel}) are solved starting from a
zero congestion state: $q_{i}^{0}=0$ $\forall i$. The evolution of the
system is monitored by means of the following order parameter
\cite{cong1}:
\begin{equation}
\rho(t)=\lim_{T\rightarrow\infty}\frac{Q(t+T)-Q(t)}{pT}\;,
\end{equation}
where $Q(t)$ is the sum of all the queue lengths at time step $t$,
$Q(t)=\sum_{i=1}^{N}q_{i}^t$. The stationary value, $\rho$, of the
above order parameter is bounded ($0\leq \rho\leq 1$) and describes
the dynamical regime in which the system ends up. Namely, $\rho=0$
indicates that the system is able to balance the incoming flow of new
packets with a successful delivery of the old ones. In this case the
system is said to operate in the {\em free-flow regime}. Instead, when
$\rho>0$ the above balance is not fulfilled and the queues of the
nodes increase their size in time at a rate $\rho\cdot p$. In this
latter situation the system is in the {\em congested phase}.

We have studied the behavior of the order parameter $\rho$ by taking
the rate of packet creation $p$ as the control parameter. The
arrival-to-destination probability is set to $\mu=0.2$ as the usual
value found in the Internet \cite{bookvesp}. The corresponding phase
diagrams are shown in Fig.~\ref{fig:1} for several values of the
rejection probability $\eta$ using a SF network of $N=5000$ with
$P(k)\sim k^{-2.2}$. As observed in the figure, the transition from
free-flow to congestion occurs in a smooth way at low values of $p$
being the critical point $p_c=0.02$ for $\eta=0$ (no
rejection). However, as the rejection rate $\eta$ increases the value
of $p_c$ decreases and $\rho$ increases faster (see inset in
Fig.~\ref{fig:1}).


\section{Analytical approximation of global congestion minimization}
\label{sec:theo}

\begin{figure}[t!]
\epsfig{file=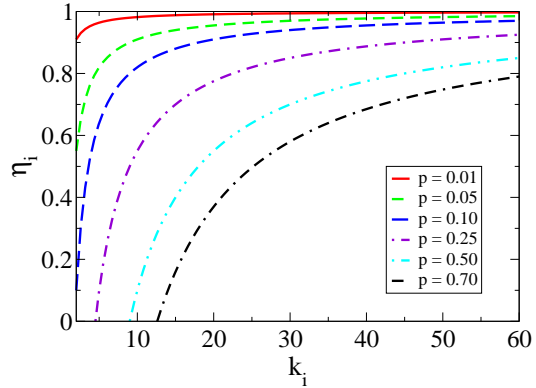,width=3.0in,angle=-0,clip=1}
\caption{(Color online). Rejection rates of nodes as a function of their degree
  $\eta_{i}(k_{i})$ as obtained from equation (\ref{eta}). The curves
  correspond to different values of the external load of information
  $p$.}
\label{fig:new}
\end{figure}

The above results question the convenience of implementing a rejection
mechanism in routing models. However, the bad performance of this
rejection mechanism relies on the homogeneous distribution of the
rejection rates across the routers of the network.  We now explore the
general situation in which the individual rejection rates are
independent. Therefore the set of equations (\ref{nullmodel})
transforms into:
\begin{eqnarray}
q_{i}^{t+1}=q_{i}^t+p&+&\sum_{j=1}^{N}\frac{\Theta(q_j^t)A_{ji}}{k_{j}}(1-\mu)(1-\eta_i)\nonumber
\\ &-&\Theta(q_{i}^t)\sum_{j=1}^{N}\frac{A_{ij}}{k_{i}}(1-\eta_j)\;.
\label{nullmodel2}
\end{eqnarray}
This new set of equations is now used to determine the optimal set
$\{\eta_{i}\}$ so that congestion is minimized for a given value of
$p$. To this aim, we first use two assumptions: {\em (i)} the nodes
have reached a stationary state, $q_{i}^{t+1}=q_{i}^t$ $\forall i$,
and {\em (ii)} the queue length of nodes is nonzero,
$\Theta(q_{i}^t)=1$ $\forall i$. These provisos admitted, equations
(\ref{nullmodel2}) turn into the following set of equations for the
rejection rates of the routers $\{\eta_{i}\}$:
\begin{equation}
0=p+\sum_{j=1}^{N}\frac{A_{ji}}{k_{j}}(1-\mu)(1-\eta)
-\sum_{j=1}^{N}\frac{A_{ij}}{k_{i}}(1-\eta)\;.
\label{stationary}
\end{equation}
Now we make use of the annealed approximation of the adjacency matrix
\cite{Bianconi1,Bianconi2,Guerra}:
\begin{equation}
A_{ij}=A_{ji}=\frac{k_{i}k_{j}}{N\langle k\rangle}\;,
\label{annealed}
\end{equation}
where $\langle k\rangle$ is the average degree of the network
($\langle k\rangle\simeq 4$ in our case). Introducing the annealed
expression (\ref{annealed}) into equations (\ref{stationary}) we
obtain:
\begin{equation}
k_{i}(1-\eta_{i})=\frac{1}{1-\mu}\left[\langle k(1-\eta)\rangle-p\langle k\rangle\right]\;,
\label{equation}
\end{equation}
where $\langle k(1-\eta)\rangle=\sum_{j}k_{j}(1-\eta_{j})/N$. Equation
(\ref{equation}) clearly shows that the larger the degree of a router
the larger its rejection rate. Therefore, from this expression we
observe that a non-homogeneous distribution of rejection rates across
the routers is beneficial to assure the free-flow condition (and thus
to delay the onset of congestion). We can calculate the expression of
the rejection rate by computing the value of $\langle
k(1-\eta)\rangle$.  From equation (\ref{equation}) we obtain:
\begin{equation}
\langle k(1-\eta)\rangle=\frac{1}{1-\mu}\left[\langle k(1-\eta)\rangle-p\langle k\rangle\right]\;,
\end{equation}
and finally we have:
\begin{equation}
\langle k(1-\eta)\rangle=\frac{p}{\mu}\langle k\rangle\;.
\end{equation}
Therefore, the rejection rate of a node with connectivity $k_{i}$ reads:
\begin{equation}
\eta_{i}=1-\frac{p\langle k\rangle}{\mu k_{i}}\;.
\label{eta}
\end{equation}
As anticipated above, expression (\ref{eta}) shows that the rejection
rates of nodes should depend on their degrees rather than being
externally set to a constant value. In Fig. \ref{fig:new} we apply
equation (\ref{eta}) to plot the rejection patterns corresponding to
different values of the external load $p$. As shown, $\eta_{i}$
decreases with $p$ and increases with $k_{i}$.

The assumptions made in order to obtain equation (\ref{eta}) point out
that its validity, for all the nodes, should be restricted to the
proximity of the critical point $p_c$. First, for $p<p_c$ many of the
queues are zero (invalidating assumption ({\em ii})) thus making the
rejection rate imposed by equation (\ref{eta}) too restrictive for the
real traffic conditions.  On the other hand, for $p>p_c$ assumption
({\em i}) does not hold for all the nodes. This is manifested by the
prediction of negative rejection rates, $\eta_{i}<0$, in equation
(\ref{eta}) for those nodes with low connectivity. In practice, the
impossibility of displaying negative rejection rates fixes their
rejection rate to $\eta_{i}=0$. However, those nodes with large enough
connectivity can still avoid congestion by means of positive rejection
rates as described in equation (\ref{eta}) (see
Fig. \ref{fig:new}). Following these arguments, we can estimate the
exact value of $p_c$ as the maximum value of $p$ for which
$\eta_{i}\geq0$ for all the nodes in the network. In particular, given
that, for a given $p$, the value of $\eta_{i}$ increases with $k_{i}$
we obtain $p_c$ imposing in equation (\ref{eta}) that those nodes with
the minimum connectivity, $k_{i}=k_{min}$, have $\eta_{i}=0$. Since in
our case $k_{min}=2$ and $\langle k\rangle\simeq 4$ we obtain
$p_c\simeq 0.1$. Therefore, by externally fixing the rejection rate of
each node as dictated by equation (\ref{eta}) we can assure the
permanence in the free-flow phase up to $p_c\simeq 0.1$.


\section{Myopic adaptability} 
\label{sec:myopic}

The minimal traffic model introduced in section \ref{sec:minimal}
shows that system's performance deteriorates as soon as rejection
rates are uniformly set in the system. However, in section
\ref{sec:theo} we have shown that a non-uniform configuration for the
rejection rates shifts the critical load to larger values. However,
this non-uniform configuration has been externally imposed and derived
analytically following different assumptions. A correct derivation of
the optimal configuration would imply, on one hand, a more
sophisticated calculation and, on the other hand, a complete knowledge
of the architecture of the network. This latter condition makes
unfeasible the external tuning of the individual rejection rates.

In order to overcome the need of global knowledge about the topology
of the network we now introduce an adaptive scheme based solely on the
local information available to nodes. In this adaptive setting we
will allow nodes to choose their own rejection rate so that the
dynamical state of a node will be described by both $q_{i}^t$ and
$\eta_{i}^t$:
\begin{eqnarray}
q_{i}^{t+1}=q_{i}^t&+&p+\sum_{j=1}^{N}\frac{\Theta(q_{j}^t)A_{ji}}{k_{j}}(1-\mu)\left[1-\eta_{i}^t\right]\nonumber
\\ &-&\Theta(q_{i}^t)\sum_{j=1}^{N}\frac{A_{ij}}{k_{i}}\left[1-\eta_{j}^t\right]
\label{adapmodel}
\end{eqnarray}
The individual choice of each instant value $\eta_{i}^t$ aims at
operating at the optimal regime as given by the external parameters
$p$ and $\mu$. To this aim, each node chooses its own rejection rate
for the following time-step attempting to reach an optimal queue
length, $q^{opt}=p/\mu$, so that traffic is homogeneously distributed
across the network. To this end, a node raises or decreases its own
rejection rate depending on the deviation of its instant queue length
from the optimal queue, $\Delta_{i}^t=q_{i}^{t}-q^{opt}$. This
rationale mimics a myopic behavior by which, regardless of the
congestion state of the system, nodes are allowed to close the door to
new packets while decreasing their respective queues. To incorporate
this adaptive behavior we couple equations (\ref{adapmodel}) with the
following evolution equations for the set $\{\eta_{i}^{t}\}$:
\begin{equation}
\eta_{i}^{t+1}=\frac{1}{1+\exp{\left(-\beta\Delta_{i}^{t}\right)}}\;.
\label{modelA}
\end{equation}
This evolution rule takes the form of the saturated Fermi function so
that congested nodes, $q_{i}>q^{opt}$, will tend to total rejection,
$\eta_{i}^{t+1}\rightarrow 1$, whereas those under-congested will open
the door to new packets, $\eta_{i}^{t+1}\rightarrow 0$. The velocity
of the transition from these two regimes is controlled by $\beta$
since it accounts for the reactivity of nodes to congestion.  Note,
that $\eta^{t+1}_{i}=0.5$ will be adopted whenever $q^{t}_i=q^{opt}$.

\begin{figure}[t!]
\epsfig{file=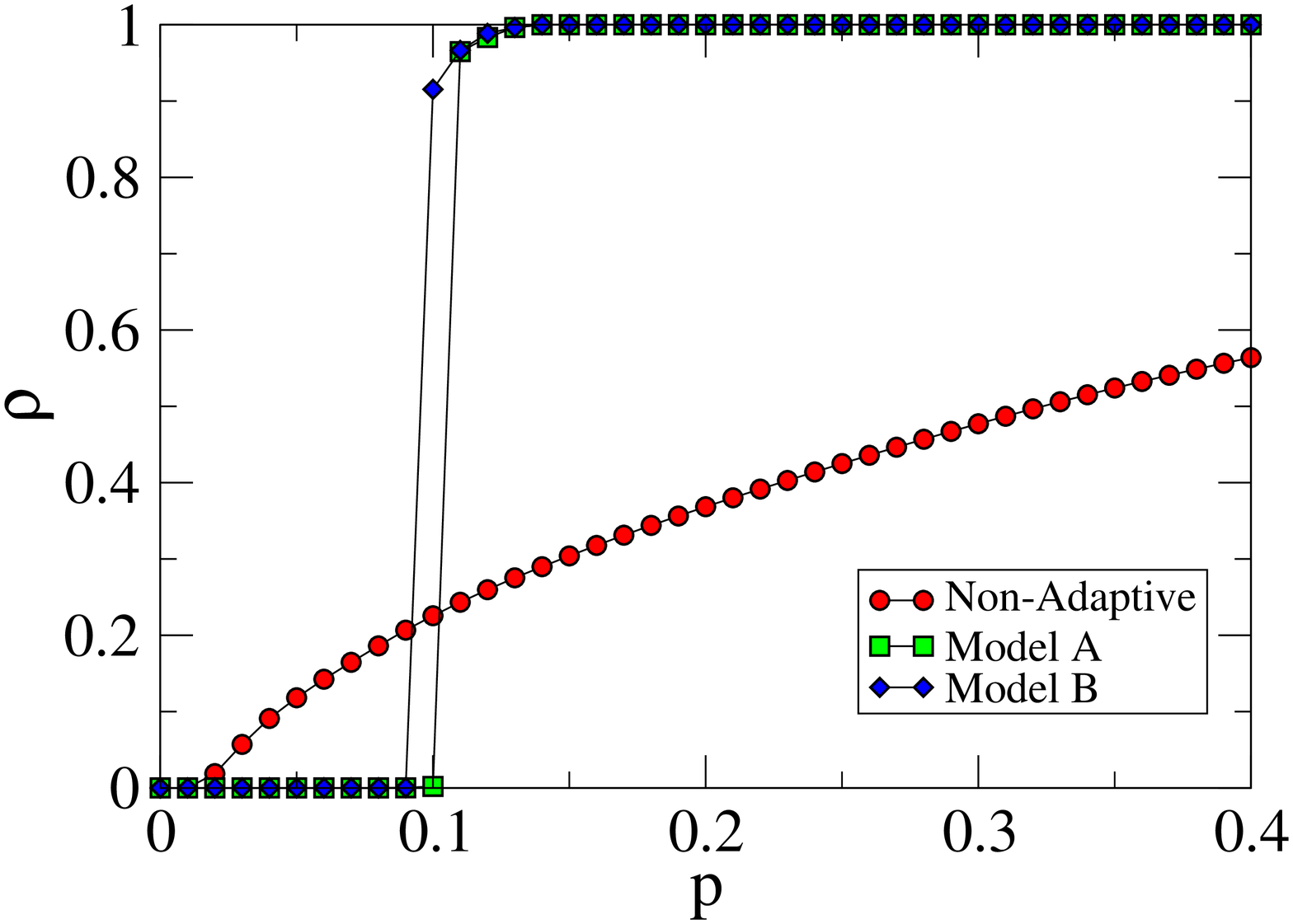,width=3.1in,angle=0,clip=1}
\epsfig{file=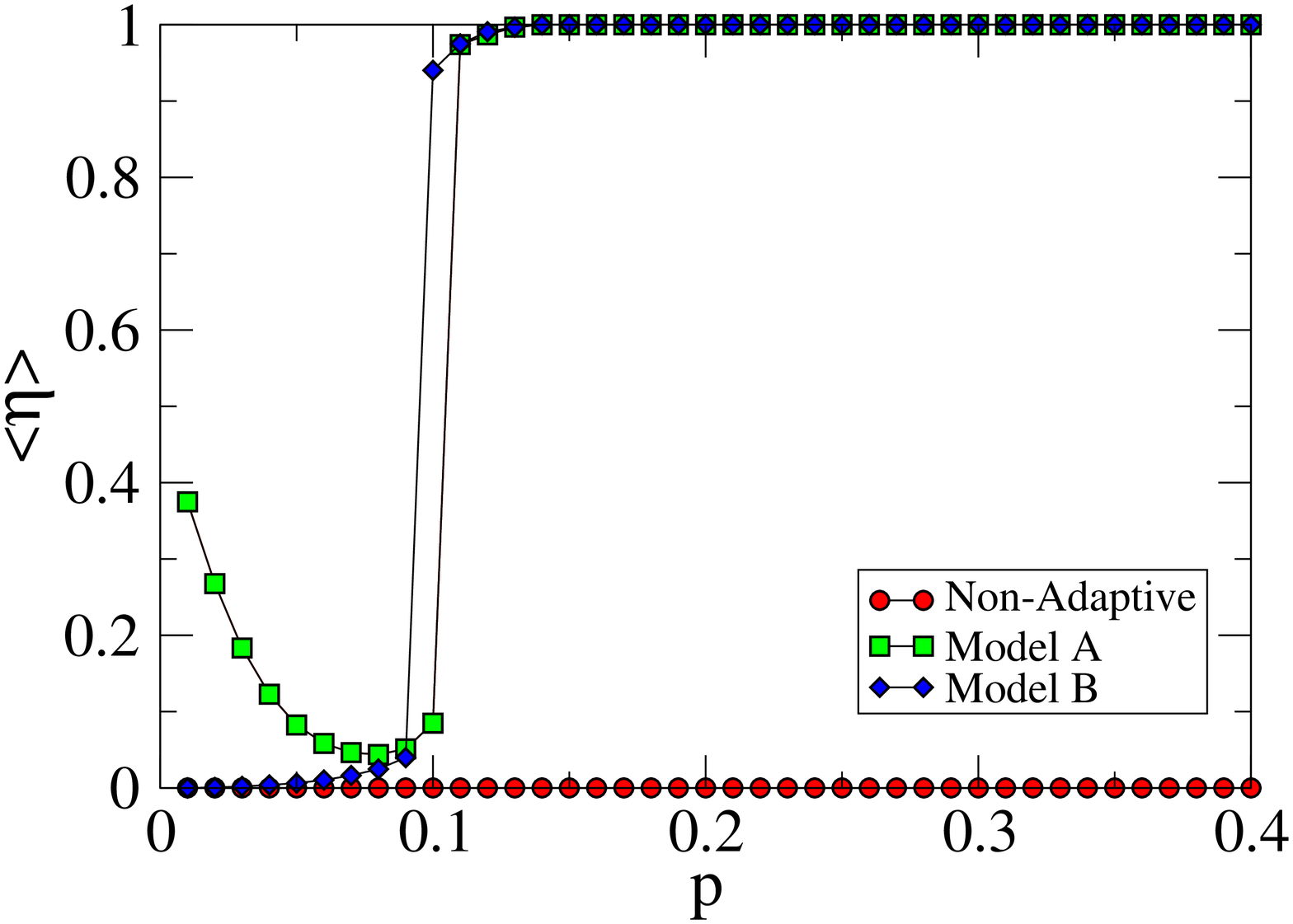,width=3.1in,angle=0,clip=1}
\caption{(Color online). (Top) Phase diagram $\rho(p)$ for the myopic
  routing models A (squares) and B (diamonds) and for the minimal
  routing model (circles). (Bottom) Average rejection rate $\langle
  \eta\rangle$ as a function of $p$ of the former three routing
  schemes.}
\label{fig:2}
\end{figure}

The adaptive equations (\ref{modelA}) allow for abrupt changes in the
rejection rates between two consecutive time steps. Thus, we also
explore a different formulation:
\begin{equation}
\eta_{i}^{t+1}=\eta_{i}^t+\beta\Delta_{i}^{t}\;,
\label{modelB}
\end{equation}
in which the rejection rates evolve smoothly. Rule (\ref{modelB}) is
completed by assuring that $\eta_{i}$ remains bounded so that $0\leq
\eta_{i}\leq 1$. In the above equation (\ref{modelB}), $\beta$ acts as
the inverse of the time between two consecutive time steps of the
adaptive dynamics. Therefore, in the continuous time approximation of
equation (\ref{modelB}), the derivative of the rejection rate is equal
to the difference between the instant queue length and its optimal
value, {\em i.e.}  $\Delta_{i}^t=q_{i}^{t}-q^{opt}$. Note that in this
setting when $q^t_{i}=q^{opt}$ a router will adopt $\eta_i^{t+1}=0$.

\begin{figure}[t!]
\epsfig{file=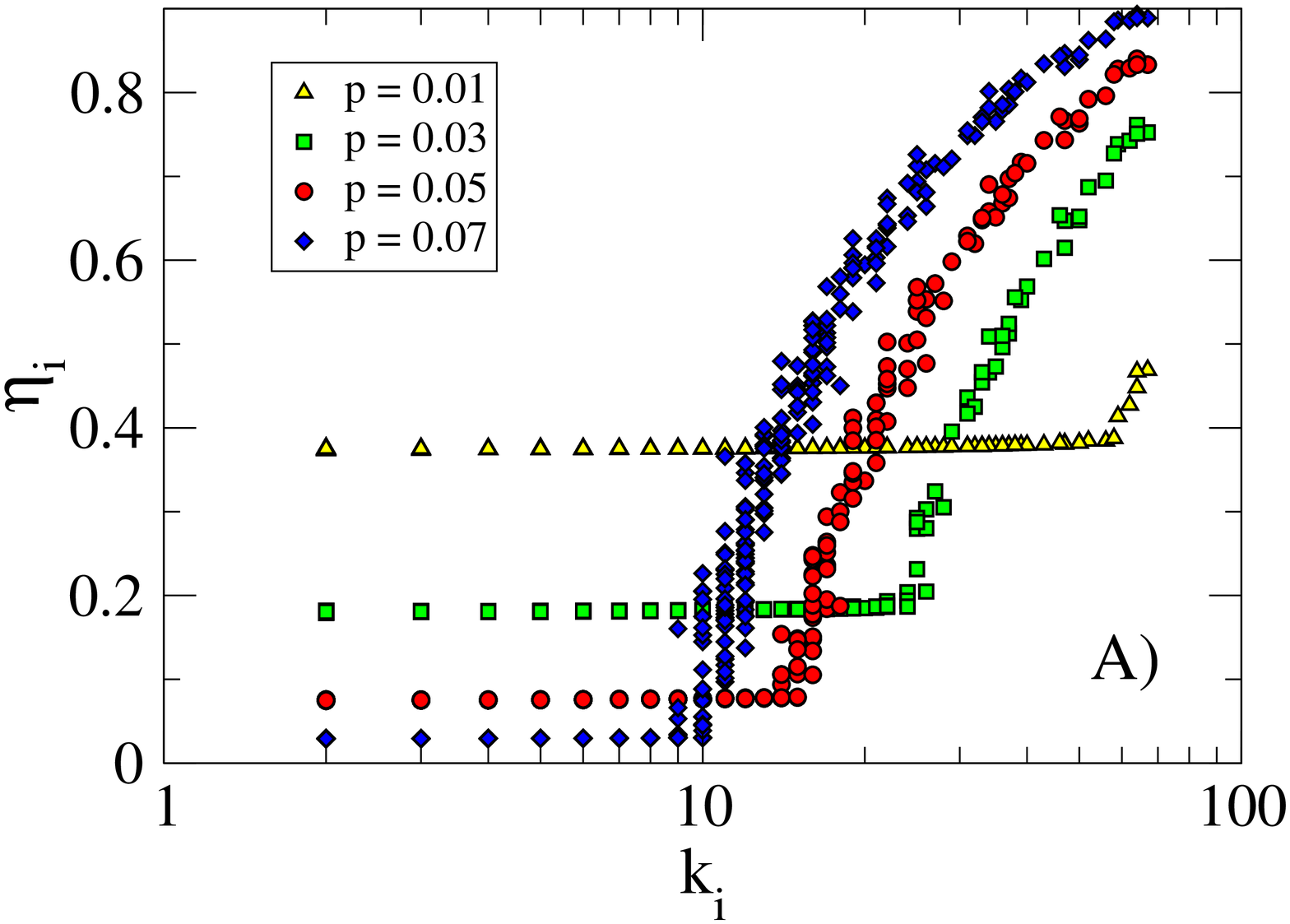,width=3.1in,angle=0,clip=1}
\epsfig{file=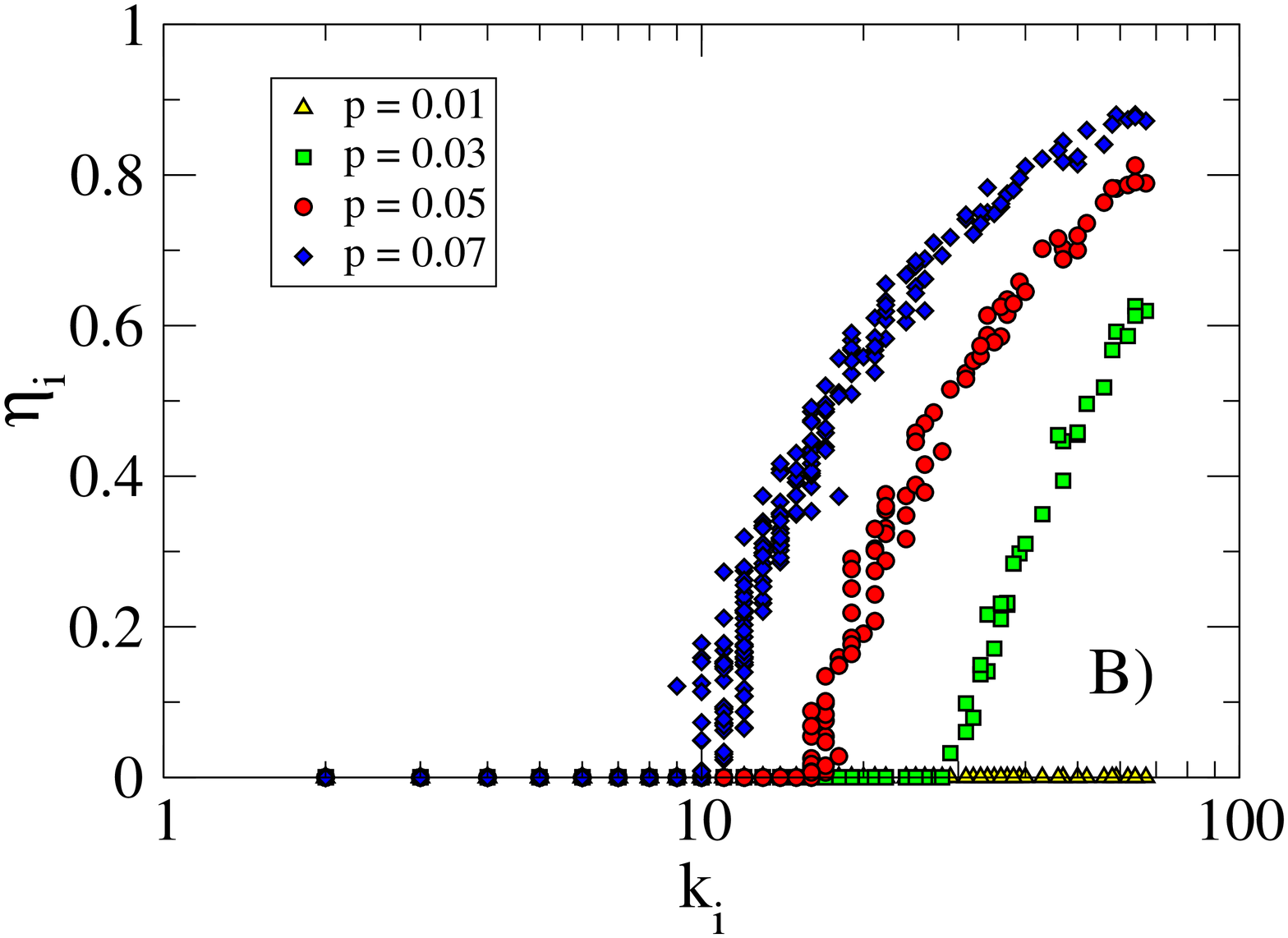,width=3.1in,angle=0,clip=1}
\caption{(Color online). Distribution of the individual rejection
  rates $\eta_{i}$ across degree-classes for several values of $p$ in
  the myopic routing models A (top) and B (bottom).}
\label{fig:3}
\end{figure}

In the following we will use the two formulations for the myopic
adaptive model and show that the results are qualitatively the
same. Namely, we will call model A to equations (\ref{adapmodel}) and
(\ref{modelA}), and model B to the formulation using equations
({\ref{adapmodel}) and ({\ref{modelB}).  Note that in both models the
    parameter $\beta$ controls  the reaction speed of
      nodes to congestion. In this direction, our numerics have shown
      that by changing $\beta$ one basically controls the duration of
      the transient time before the stationary distribution of the
      rejection rates is reached. In the following, we set $\beta=10$
    and $\beta=10^{-2}$ in models A and B respectively.

In the top panel of Fig.~\ref{fig:2} we show the phase diagram,
$\rho(p)$, of the myopic adaptive model with the two formulations. As
observed, in both formulations the myopic model displays an abrupt,
first-order like, transition from the free-flow to the congested
state. Moreover, in Fig.~\ref{fig:2} we have also plotted the phase
diagram of the minimal model when $\eta=0$, {\em i.e.} its most
congestion-resilient version, to show the improvement of myopic
adaptability by shifting the jamming transition from $p_c=0.02$ to
$p_c\simeq 0.1$. This value for the critical load is exactly the same
as the one predicted in section \ref{sec:theo} by means of the
analytical approximation using global knowledge.  Thus, the myopic
adaptive model, equals the delay of the congestion onset obtained by
minimizing congestion globally.

To analyze the roots of the resilience of the myopic adaptive routing
to congestion we have plotted in the bottom panel of Fig. \ref{fig:2}
the mean value of the rejection rate, $\langle
\eta\rangle=\sum_{i=1}^{N}\eta_{i}$. In this case we observe that
models A and B display the same pattern after the sharp transition to
congestion, {\em i.e.} the sudden closing of all the doors in the
network thus causing the abrupt transition to $\rho\simeq 1$ as soon
as $p>p_c$. On the other hand, the configurations adopted by both
models before the onset of congestion, $p<p_c$, are quite different:
While in model B $\langle \eta\rangle\simeq 0$, for model A a
significant part of the population adopts $\eta_{i}>0$. Surprisingly,
in this latter setting the average rejection rate decreases as we
approach the critical point, $p_c$.

To have a deeper insight about the microscopic configurations that
allow to delay the onset of congestion we show in Fig. \ref{fig:3} the
set of individual rejection rates of nodes $\{\eta_{i}\}$ ranked
according to their degrees. In both models A and B, the correlation
between $\eta_{i}$ and $k_{i}$ is clear since all the routers within
the same degree-class display similar rejection rates. First, in model
A we observe that for $p=0.01$ the system self-organizes homogeneously
around $\eta\simeq0.4$. However, when $p$ increases the rejection
rates of low-degree classes decreases while hubs start to close their
doors progressively as $p$ increases. For model B the microscopic
configurations adopted as $p$ increases are similar regarding the
behavior of high-degree nodes. However, in this latter scenario
low-degree nodes remain accepting incoming packets up to the congested
state. These two figures show that the two different internal dynamics
(showing different microscopic organizations ) lead to the same
macroscopic result: the delay of the onset of congestion.

Let us highlight that the delay of the congestion onset in this myopic
adaptive setting again contradicts the results obtained for the
minimal routing model in which, even a small (homogeneously
distributed across routers) rejection rate leads to an increase of the
congestion in the system. Quite on the contrary, the myopic adaptive
model points out the same idea concluded from the global minimization
of congestion: a hierarchical (degree-based) organization of the
rejection rates by the system is strongly beneficial to avoid the
congestion of the system. However, from figure \ref{fig:3} it becomes
evident that the strategies self-adopted in the myopic adaptive
settings are clearly different than the ones obtained in section
\ref{sec:theo} from equation (\ref{eta}) when congestion was minimized
using global knowledge. Although in equation (\ref{eta}) the value of
the rejection rate increases with the degree of the node (as in the
myopic setting), the evolution with $p$ is quite different. Thus,
although the critical load has been shifted to the same value as the
one found in section \ref{sec:theo}, the self-organized patterns of
the rejection rates in the myopic settings reveal a clearly different
scenario.

\section{Empathetic adaptability} 
\label{sec:empathetic}

\begin{figure}[t]
\epsfig{file=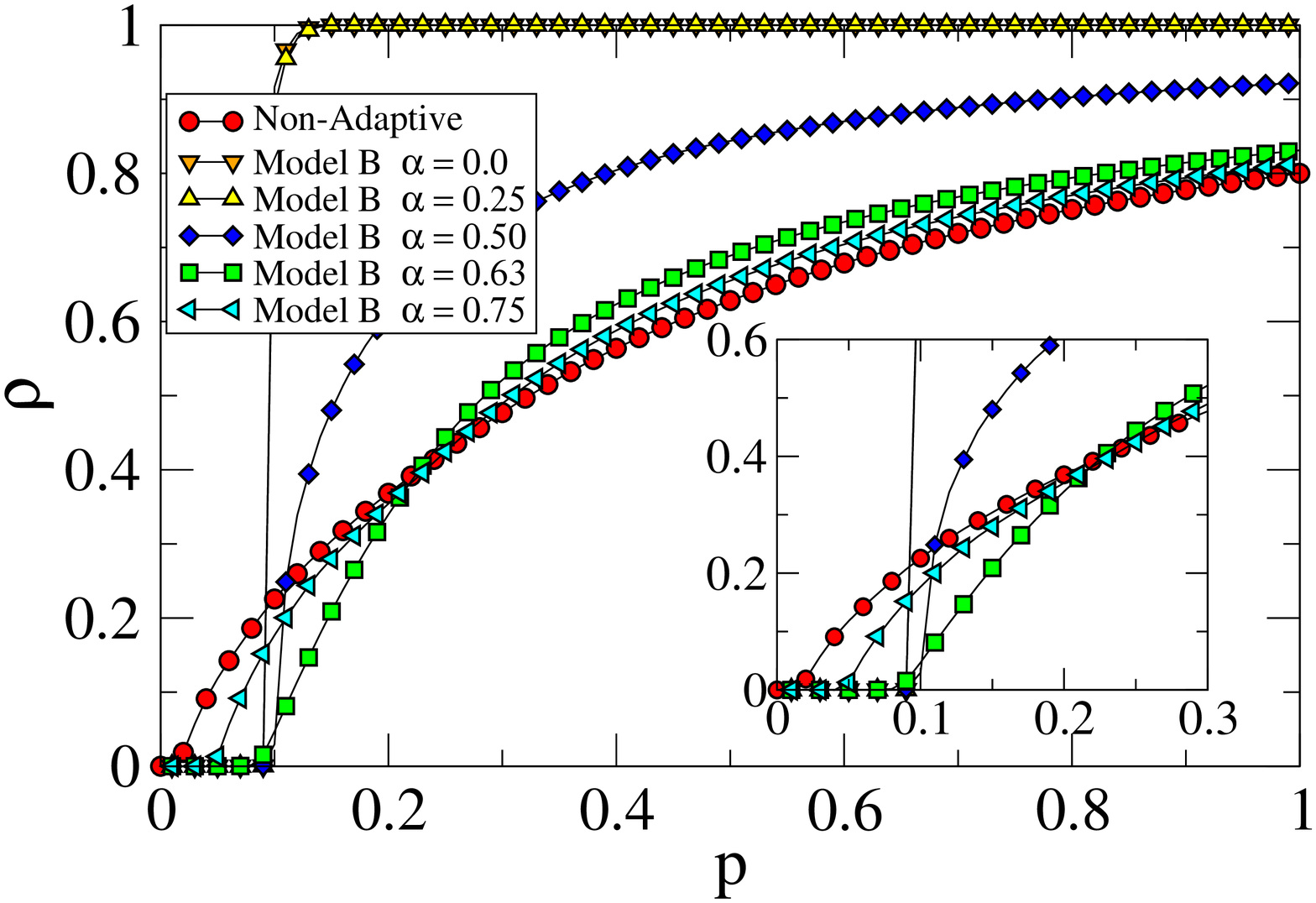,width=3.1in,angle=-0,clip=1}
\epsfig{file=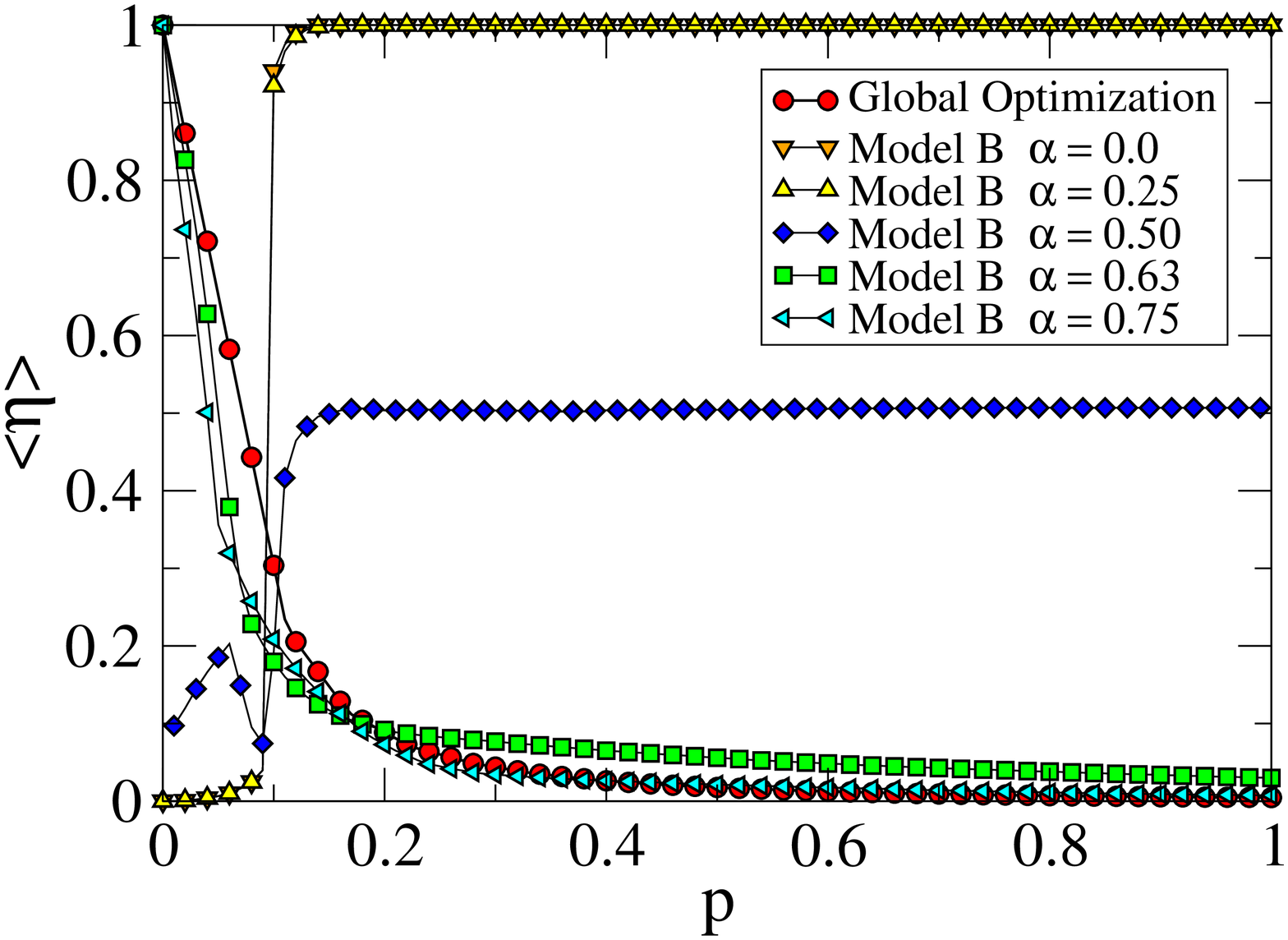,width=3.1in,angle=-0,clip=1}
\caption{(Color online). (Top) Phase diagram $\rho(p)$ of the
  empathetic routing model for several values of the empathy parameter
  $\alpha$. The phase diagram of the minimal routing model (circles)
  is also plotted for the sake of comparison. (Bottom) For the same
  empathy parameters we show the average rejection rate $\langle
  \eta\rangle$ as a function of $p$. The function $\langle eta\rangle
  (p)$ obtained analytically from global minimization and computed
  from equation (\ref{eta}) is also shown.}
\label{fig:4}
\end{figure}

The myopic adaptive setting has improved remarkably the resilience to
congestion without the need of tuning any external
parameters. However, the existence of an abrupt phase transition,
again as found in \cite{cong3,cong4,cong8,cong9}, demands for further
improvements. The main goal in order to soften such abrupt transition
is to avoid that all the nodes close their doors due to its own
congestion by incorporating an empathetic behavior based on the local
knowledge about the dynamical state of their neighbors. This
empathetic behavior should motivate congested nodes to open their
doors when detecting an hyper-congested state in its surroundings. To
this aim we take model B \cite{note} and reformulate its equations as
follows:
\begin{equation}
  \eta_{i}^{t+1}=\eta_{i}^t+\beta\left[(1-\alpha)\Delta_{i}^{t}-\alpha\langle\Delta_{j}^{t}\rangle_{\Gamma_{i}}\right]\;.
\label{modelcoop}
\end{equation}
In the above equations we introduce a new term accounting for the
average level of congestion in the neighborhood, $\Gamma_{i}$, of a
node $i$,
\begin{equation}
 \langle\Delta_{j}^{t}\rangle_{\Gamma_{i}}=\sum_{j=1}^{N}\frac{A_{ij}}{k_{i}}\Delta_{j}^{t}\;.
\end{equation}
The relative importance that nodes assign to the local level of
congestion in their neighborhoods with respect to their own state is
controlled by the parameter $\alpha$. In particular, when $\alpha=0$
we recover the myopic setting whereas for $\alpha=1$ routers behave
``altruistically'' and their decisions are based solely on their
neighbor's state of congestion. Thus, the parameter $\alpha$ measures
the degree of empathy between connected routers.

In the top panel of Fig. \ref{fig:4} we plot the phase diagrams for
several values of $\alpha$ together with that of the minimal
non-adaptive routing model. We observe that for $\alpha<0.5$ the
phase-transition is similar to that of the myopic adaptive model
($\alpha=0$), {\em i.e.}  showing a critical load of $p_c\simeq 0.1$
followed by a first-order transition to full congestion. However, from
the figure we observe that when $\alpha>0.5$ the transition to
congestion occurs smoothly, thus recovering the behavior of the
minimal model. On the other hand, the value of $p_c$ also decreases
with $\alpha$ (thus anticipating the onset of congestion) although it
remains close to the original value $p\simeq 0.1$ until $\alpha\simeq
0.63$. Moreover, for $p>p_c$, the curves corresponding to
$\alpha=0.63$ and $\alpha=0.75$ reach levels of congestion similar to
those observed in the minimal model.

\begin{figure*}[t]
\epsfig{file=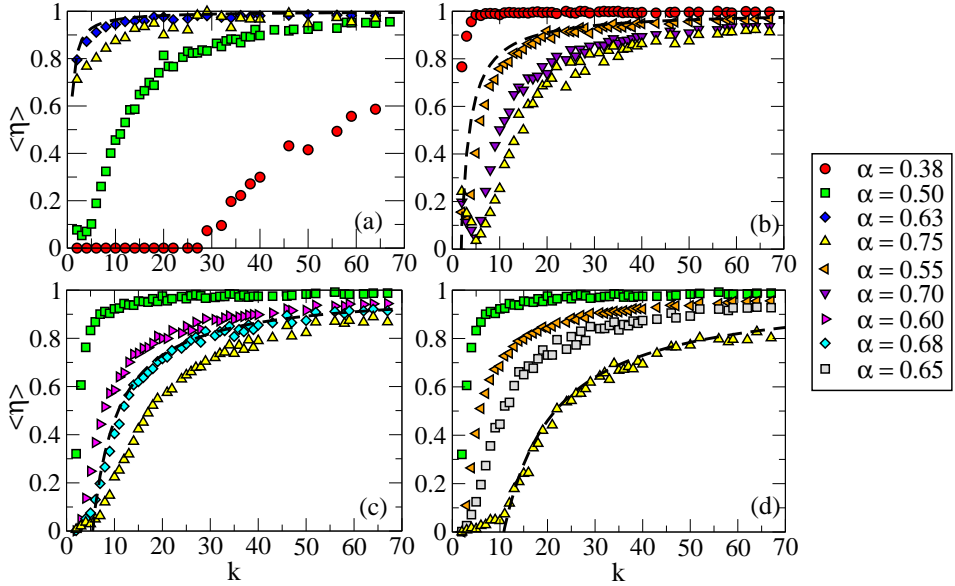,width=5.2in,angle=0,clip=1}
\caption{(Color online). Distribution of the mean rejection rates
  $\langle \eta \rangle$ across degree-classes of the empathetic
  adaptive model for several values of the empathy parameter $\alpha$
  compared with the global minimization prediction (dashed
  line). Different traffic values $p$ are presented: (a) $p = 0.02$
  (free-flow regime), (b) $p = 0.1$ (critical point), (c) $p=0.3$ and
  (d) $p= 0.6$ (congested state)}
\label{fig:5}
\end{figure*}

In order to gain more insight about the strategy adopted in the
empathetic setting we have computed the average level of rejection
rate as a function of $p$ for the relevant values of $\alpha$. In the
bottom panel of Fig. \ref{fig:4} we observe that those curves
corresponding to $\alpha>0.5$ are quite different from those obtained
in Fig. \ref{fig:2} for the myopic adaptive setting. In particular,
when $p\ll p_c$ the empathetic adaptability shows a large amount of
rejection. However, as $p$ increases the average rejection rate
decreases monotonously. This high-rejecting behavior for $p<p_c$, was
not observed in the myopic scheme. Quite on the contrary, it was shown
that nearly all the doors were open in the sub-critical
regime. However, the high rate of rejection observed in
Fig. \ref{fig:4} is due to the large degree of empathy ($\alpha>0.5$)
and the existence of a number of under-congested nodes,
$\Delta_{i}<0$, in the sub-critical regime. Under these low traffic
conditions, most nodes will close partially their doors when detecting
under-congested neighborhoods,
$\langle\Delta_{j}^{t}\rangle_{\Gamma_{i}}<0$, in order to benefit
from the availability of neighbors to handle their packets. This
situation is highly dynamical and most of the nodes experiment large
fluctuations in their rejection rates until the system equilibrium is
reached. This microscopic scenario, although clearly different from
that of the myopic setting, enables to delay the onset of congestion
in an efficient way. On the other hand, as $p$ approaches $p_c$ and
for $p>p_c$ we observe that (for $\alpha>0.5$) the value of $\langle
\eta\rangle$ decreases to $0$ as $p$ increases. This is due to both
the large number of over-congested neighborhoods,
$\langle\Delta_{j}^{t}\rangle_{\Gamma_{i}}>0$, surrounding routers in
the super-critical regime and their large degree of empathy. As
expected, empathy prevents from the sudden door closing when $p>p_c$,
thus favoring a smooth phase transition displaying congestion levels
simlar to those observed in the minimal routing model in the
super-critical regime.

Interestingly, the monotonous decrease of $\langle \eta\rangle(p)$
from $\langle \eta\rangle=1$ at $p=0$ shown in Fig. \ref{fig:4},
points out a similar behavior to that obtained by means of global
minimization of congestion. As shown in the bottom panel of Fig.
\ref{fig:4} the theoretical estimation of $\langle \eta\rangle(p)$
(circles) follows the same trend as the self-adopted patterns for
$\alpha>0.5$. To analyze in detail the similarity between the
empathetic setting and the microscopic patterns predicted by global
minimization of congestion we plot in Fig. \ref{fig:5} the average
value of the rejection rate as a function of the degree $k$ of the
nodes, $\langle \eta\rangle(k)$, for several values of $p$ and
$\alpha$. The panels correspond to (a) $p=0.02$ (free-flow regime),
(b) $p=0.1$ (critical point), (c) $p=0.3$ and (d) $p=0.6$ (congested
state). The shape of each curve $\langle \eta\rangle(k)$ behaves
similarly to the theoretical one as predicted from equation
(\ref{eta}). More importantly, for each value of $p$ there is one
value of $\alpha$, $\alpha^{opt}$, for which the curve $\langle
\eta\rangle(k)$ fits perfectly the prediction made by global
minimization of congestion. The precise value of $\alpha^{opt}$
depends on $p$. In particular, for $p=0.02$ we find
$\alpha^{opt}\simeq 0.63$, for $p=0.1$ we obtain $\alpha^{opt}\simeq
0.55$, for $p=0.3$ we have $\alpha^{opt}\simeq 0.68$ and, finally, for
$p=0.6$ the value found is $\alpha^{opt}\simeq0.75$. Moreover, from
the top panel of Fig. \ref{fig:4}, we observe that the values found
for $\alpha^{opt}$ are those for which congestion, $\rho(p)$, is
minimum. This result points out that empathetic adaptability is able
to avoid congestion by means of only local information as much as
global minimization does.


\section{Conclusions}

We have studied a novel mechanism that allows routers to adapt their
individual strategies based on their local knowledge about
congestion. Although in our approach nodes can only decide either to
refuse or to accept incoming packets from their first neighbors, we
obtain a variety of dynamical behaviors. First, we have analyzed the
situation when no individual adaptability is allowed. This allows us to
show that whenever a small level of rejection is applied indistinctly
to all the nodes, one obtains a worse overall behavior than when all
incoming flows are accepted by the routers. Then, we have considered
that routers can have different rejection rates and derived
analytically their patterns to minimize congestion, considering global
knowledge of the network topology. With these globally optimized
patterns the resilience to congestion of the system can be enhanced
significantly. Besides, these patterns reveal a dependence of the
rejection rate and the degree of the router while its mean value decays
with the incoming load of packets.

After deriving global minimization of congestion we have studied the
situation in which nodes self-adjust their own rejection rates
dynamically depending on their instant level of congestion (myopic
setting). In this case we have shown that the critical load of the
network is shifted to a value similar to that found analytically by
means of global minimization of congestion. This improvement is again
achieved by a proper distribution of the rejection rates according to
the degrees of the routers. However, in the adaptive case, such
degree-correlated configuration is self-tuned by the system and
differs from that obtained analytically. As usual in congestion-aware
routing schemes, such delay in the congestion onset comes together
with an abrupt transition from the free-flow phase to the congested
one that prevents from having any warnings of the approach to the
onset of congestion. For this reason, we have finally explored the
situation in which routers also consider the congestion state of their
first neighbors to adapt their rejection rates. We have shown that
when nodes empathize with the congestion state of their neighbors,
thus not rejecting packets from them when they detect an
over-congested neighborhood, the shift in the critical load (obtained
through global minimization and the myopic adaptability) is preserved
and followed by a smooth congestion transition. Moreover, the analysis
of the microscopic patterns of rejection rates when empathy is the
mechanism at work points out a similar organization to that obtained
from global minimization. In particular, it is possible to find the
degree of empathy that perfectly agrees with the analytical estimation
of the rejection pattern that minimize congestion for a given load of
information.

In summary, we have shown that allowing routers to adapt their own
strategies together with a certain degree of local empathy is strongly
beneficial to the behavior of complex communication systems. Moreover,
the improvement shown when local empathy is at work is similar to that
obtained by minimizing congestion by means of a global knowledge of
the network topology.  Thus, the empathetic setting represents, a
remarkable example of how local rules can achieve levels of
functioning as optimal as those obtained with global knowledge of the
system. Besides, our results open the relevant question about how
local empathy can be naturally tuned as a function of the external
inputs.


\begin{acknowledgments}
This work has been supported by the Spanish Ministry of Science and
Innovation (MICINN) through Grants FIS2008-01240 and MTM2009-13848.
We acknowledge the hospitality of the members of the ATP group at the
Universit\'a di Catania where part of this work was developed. We are
grateful with Y. Moreno for his useful suggestions. We thank
A. Iniesta {\em et al.} for providing with inspiration.
\end{acknowledgments}


\begin{thebibliography}{99}
\bibitem{note} The generalization of Model A could be done in the same way.
\end{thebibliography}
\end{document}